\documentclass[11pt,twoside]{article}

\def\simlt{\mathrel{\hbox{\rlap{\hbox{\lower4pt\hbox{$\sim$}}}\hbox{$<$}}}}
\def\simgt{\mathrel{\hbox{\rlap{\hbox{\lower4pt\hbox{$\sim$}}}\hbox{$>$}}}}
\def\simlt{\mathrel{\hbox{\rlap{\hbox{\lower4pt\hbox{$\sim$}}}\hbox{$<$}}}}
\def\simgt{\mathrel{\hbox{\rlap{\hbox{\lower4pt\hbox{$\sim$}}}\hbox{$>$}}}}
\def\lesssim{\mathrel{\hbox{\rlap{\hbox{\lower4pt\hbox{$\sim$}}}\hbox{$<$}}}}
\def\gtrsim{\mathrel{\hbox{\rlap{\hbox{\lower4pt\hbox{$\sim$}}}\hbox{$>$}}}}

\usepackage{asp2006}
\usepackage{epsf}
\usepackage{psfig}
\usepackage{lscape}

\begin{document}
\title{Gamma-Ray Burst Central Engines: Black Hole Vs. Magnetar}   %%% Fill in title
\author{Brian D.~Metzger\altaffilmark{1}}   %%% Fill in author names
\affil{\center{Department of Astrophysical Sciences, Princeton University,\,\,\,\,\,\,\,\,\,\,\,\,\,\, Princeton, NJ 08544 USA}}    %%% Fill in author affiliations
\altaffiltext{1}{NASA Einstein Fellow}

\begin{abstract}{Discovered over forty years ago, Gamma-Ray Bursts (GRBs) remain a forefront topic in modern astrophysics.  Perhaps the most fundamental question associated with GRBs is the nature of the astrophysical agent (or agents) that ultimately powers them: the central engine.  In this review, I focus on the possible central engines of long-duration GRBs, and the constraints that present observations place on these models.  Long GRBs are definitively associated with the deaths of massive stars, but whether the central engine is an accreting black hole or a rapidly-spinning, highly-magnetized neutron star (a ``proto-magnetar'') remains unsettled.  This distinction has been brought into particular focus by recent MHD simulations of the core-collapse of massive, rotating ``collapsar progenitors,'' which suggest that powerful magneto-centrifugal outflows from the proto-neutron star may stave off black hole formation entirely.  Although both black hole and magnetar GRB models remain viable, I argue that the magnetar model is more mature in the sense that it provides {\it quantitative} explanations for the durations, energies, Lorentz factors, and collimation of long GRB outflows.  Given these virtues, one promising strategy to break the present stalemate is to further develop the magnetar model until inescapable (and falsifiable) predictions emerge.  This course of action signals a renewed challenge to translate time-dependent jet properties (power, magnetization, and Lorentz factor) into observables (gamma-ray light curves and spectra).}
\end{abstract}

\section{Introduction}

Despite being discovered over forty years ago (Klebesadel et al.~1973), Gamma-Ray Bursts (GRBs) largely remain enigmatic.  New observations, especially by the {\it Swift} and {\it Fermi} satellites, are continuing to advance our understanding of these events and many puzzles have been resolved.  However, as new phenomenology has been uncovered, previous assumptions have been challenged and many new questions have emerged (see, e.g., Gehrels et al.~2009 for a recent review).  Furthermore, answers to many of the {\it most fundamental} questions associated with GRBs remain (somewhat exasperatingly) elusive; for instance, ``What emission mechanism creates the gamma-rays? (e.g., synchrotron, inverse-Compton, hadronic?),'' ``What is the composition of the GRB-powering outflows? (e.g., baryons or $e^{-}/e^{+}$ pairs?),'' ``In what form is the outflow's energy stored (e.g., kinetic energy or Poynting flux?),''``How is this ordered energy `randomized' and how are the radiating charges accelerated?,'' ``How are the magnetic fields that appear necessary for the prompt and afterglow emission produced? (e.g., are they generated by shocks or carried out from the outflow's source?).''   

Perhaps the key question associated with GRBs, and the focus of this review, is the nature of the astrophysical agent (or agents) that ultimately powers them: the ``central engine'' (CE).  Soon after their discovery, there were perhaps more theories for GRBs than theorists (Ruderman 1975).  The presently appreciated requirements of supernova-scale energies, short timescales (down to milliseconds), and relativistic speeds (Lorentz factors $\gtrsim 100$) have, however, significantly narrowed the possibilities: GRBs are almost certainly the result of stellar-mass black holes (BHs) or neutron stars (NSs) being formed or undergoing catastrophic rearrangement (e.g., Katz 1997).  All plausible CEs are powered by either accretion or rotation, which makes a deep gravitational potential and a reservoir of significant angular momentum the key ingredients of any model.  A full understanding of GRBs may ultimately require comprehending the interplay between the physics of relativistic fluid dynamics, ultra-strong gravity, strong electromagnetic fields, nuclear/weak interactions, and plasma processes such as collisionless shock formation, non-thermal particle acceleration, and magnetic reconnection.  The ways in which these nominally disparate physical processes conspire to produce a GRB makes studying the CE both exciting and uniquely challenging.  

Our discussion will focus on simple theoretical models of GRB CEs in order to elucidate both the {\it abilities} and the {\it limitations} of these prime movers.  Fundamental questions about GRBs (such as those raised in the first paragraph) are typically addressed by taking an agnostic approach towards the nature of the CE.  Pursued in the spirit of ``model independence,'' this approach also often allows the CE undue freedom, making it ``a flexible source of power whose properties are only limited by its total mass and the referees of theoretical papers'' (Blandford 2002).  It is important to emphasize that rather stringent constraints can be placed on GRB models by employing a self-consistent physical model for the CE.  It is certainly true, for instance, that the formation, collimation, and stability of ultra-relativistic jets remains a formidable unsolved theoretical problem (e.g. McKinney $\&$ Blandford 2009).  However, other characteristics of the CE (such as the mass loading of the jet and energy budget) can, in some cases, be evaluated with more confidence.  These aspects of the problem are more theoretically tractable largely because the immediate vicinity of the CE is often highly collisional and (local) kinetic equilibrium is assured; in this sense, the ``extreme'' environment of the CE can be a blessing.  

The review is organized as follows.  I begin in ($\S\ref{sec:obs}$) with a brief historical background and summary of the basic observational constraints on CE models.  It has long been known that GRBs come in (at least) two varieties (``long'' and ``short''), separated by their duration $\sim 2$ seconds (Mazets et al.~1981) and spectral ``hardness'' (Kouveliotou et al.~1993), and {\it Swift} has enabled the discovery that long and short GRBs indeed have distinct stellar progenitors (see, e.g., Berger 2009).  In this review I focus on long-duration GRBs, which are generally thought to result from the deaths of massive stars (Woosley $\&$ Bloom 2006).  For a discussion of short GRB central engines, I refer the reader to excellent reviews by Nakar (2007) and Lee $\&$ Ramirez-Ruiz (2007).  The body of the review divides into a discussion of collapsar (black hole; $\S\ref{sec:LGRBs}$) and proto-magnetar ($\S\ref{sec:magnetar}$) central engine models.  I conclude and discuss the prospects for distinguishing these models in ($\S\ref{sec:conclusions}$).  Since GRBs comprise an enormous research field, the topics covered here are necessarily limited and my reference list is only representative.  Furthermore, the discussion and emphasis necessarily reflects my own work and biases.  I apologize for any glaring oversights or omissions and refer the reader to several other excellent reviews for a different perspective: Lyutikov $\&$ Blandford (2003); Spruit (2004); Zhang $\&$ M{\'e}sz{\'a}ros (2004); Piran (2005); and Thompson (2008).

\section{Historical Development and Observational Constraints}
\label{sec:obs}

GRBs were discovered using the Vela satellites in the late 1960s, but they were not announced until 1973 (Klebesadel et al.~1973).  GRBs manifest as a deluge of high-energy emissions with characteristic durations of $\sim$ milliseconds to minutes; the light curves display rapid variability and a non-thermal (broken power-law) spectrum that peaks in the sub-MeV energy range but that often possesses a high-energy tail with significant power above $\sim$ MeV (known as a ``Band'' spectrum; see e.g. Fishman $\&$ Meegan 1995 for a review).  The fact that GRBs are variable on timescales typically $\ll$ than their total duration suggests that the outflow producing the GRB is best described as a quasi-continuous wind from the CE, rather than a singular explosion (Kobayashi et al~1997; see, however, Narayan $\&$ Kumar 2009).  A recent discovery made with {\it Fermi} is that some bright GRBs are accompanied by an additional, typically flat-spectrum GeV component (Abdo et al.~2009b), which is generally delayed with respect to the normal Band-spectrum photons by a few seconds.  While the $\lesssim$ MeV flux from GRBs normally decreases rapidly once the GRB is finished (e.g. Barniol-Duran $\&$ Kumar 2009), the GeV component appears to decay relatively gradually in time ($\propto t^{-1.5}$).   Although the origin of this emission is currently debated, it may result from the onset of the afterglow emission as the GRB-producing outflow first interacts with the surrounding, circumburst medium (e.g. Kumar $\&$ Barniol-Duran 2009).   

As cosmological sources, the implied isotropic energies of GRBs are enormous, ranging from $\sim 10^{49}$ ergs (for some short GRBs and local, sub-luminous long bursts) up to $\gtrsim 10^{54}$ ergs $\sim M_{\sun}c^{2}$ in extreme cases (e.g. the recent {\it Fermi}-detected burst GRB 080916C had $E_{\rm iso} \approx 8\times 10^{54}$ ergs; Abdo et al.~2009a).  If such a huge energy flux truly originated from a region with a size $\sim 100$ km (typical of that inferred from the observed millisecond variability), the opacity to pair-production in the high-energy photon power-law tail would be enormous (the implied optical depth for photon-photon and photon-electron interactions is $\tau_{\gamma-\gamma},\tau_{e-\gamma} \sim 10^{15}$ for typical parameters).  This requires the formation of a thermal pair photosphere, which is inconsistent with the observed non-thermal spectrum.  In order to overcome this ``compactness problem,'' the GRB-producing region must be expanding towards the observer ultra-relativistically, with a bulk Lorentz factor $\Gamma \gtrsim 100$ (e.g., Lithwick $\&$ Sari 2001).  One reason that relativistic bulk motion alters the conclusion that high-energy gamma-rays may escape is that the observer, by naively inferring the size of the emitting region from the variability, is tricked by relativistic effects into believing that the emitting region is more compact than its true physical dimensions (indeed, current models place the location of GRB emission at radii ranging from $10^{12}-10^{18}$ cm, {\it not} 100 km; compare, for instance, the emission models of Thompson 2006 and Lyutikov $\&$ Blackman 2001).  Relativistic speeds in GRB outflows have been confirmed observationally by the angular expansion rate inferred from the quenching of scintillation in a radio afterglow (Frail et al.~1997) and by direct VLBI imaging of an expanding GRB blast wave (Taylor et al.~2004).

Since GRB outflows are ultra-relativistic, observers are only in causal contact with a small solid angle ($\sim 1/\Gamma^{2}$) on the surface of the outflow (as subtended by the CE).  Thus, even if GRB outflows are significantly collimated, evidence for jet-like angular structure only becomes apparent long after the main GRB event, once the flow has slowed to a Lorentz factor $\Gamma \sim 1/\theta_{j}$, where $\theta_{j}$ is the opening angle of the jet.  Indeed, observational evidence such as ``jet breaks'' in the late-time afterglow (e.g. Rhoads 1997; Cenko et al.~2009) and the calorimetry provided by late-time radio afterglows (e.g. Berger et al.~2003) strongly suggest that GRB outflows are collimated to some degree.  This implies that although the isotropic-equivalent energies of GRBs are enormous, the true energy budget (after correcting for beaming) is much lower, typically comparable to the kinetic energy of a SN, $\sim 10^{51}$ ergs (Frail et al.~2001).  What makes the GRB CE unique is therefore not necessarily its large energy budget, but rather {\it its ability to place a significant fraction of this energy into material with ultra-relativistic speeds}.  This is an important diagnostic because it implies that the outflow remains relatively ``clean'' despite originating from what is potentially a rather dense and messy environment around the CE.  A jet with energy $\sim 10^{51}$ ergs must, for instance, entrain $\lesssim 10^{-5}M_{\sun}$ to achieve $\Gamma \gtrsim 100$.  Because all viable CE models are surrounded by significantly more mass than this ($\sim 10^{-2}-10M_{\sun}$), one of the major challenges of any model is to produce an outflow that avoids being polluted by too much of this surrounding material.

An important discovery made with ${\it Swift}$'s X-ray telescope (XRT) is that GRBs are often followed by late-time X-ray flaring on timescales of $\sim$ minutes-hours following the burst (e.g. Burrows et al.~2005).  Because of the flares' rapid evolution and other similarities to prompt GRB emission, they most likely result from continuing CE activity (e.g. Lazatti $\&$ Perna 2007).  The ``plateau'' phase observed in many X-ray afterglows on timescales $\sim 10^{3}-10^{4}$ seconds may also be powered by the CE (e.g. Panaitescu 2008; Kumar et al.~2008; see, however, Uhm $\&$ Beloborodov 2008).  Thus, an important additional constraint on CE models is the ability to be active long after the GRB itself.

\section{Collapsar (Black Hole) Model}
\label{sec:LGRBs}

The earliest cosmological CE models, such as neutron star mergers, focused (justifiably) on systems thought to naturally provide a relatively baryon-clean environment from which to launch a relativistic jet.  In 1993 Stan Woosley made the bold suggestion that GRB jets may be produced even in the dense environment of the core-collapse of a massive star (Woosley 1993).  In this ``collapsar'' model, the GRB is powered by accretion of the stellar envelope onto a BH that forms soon after the collapse (e.g. Proga $\&$ Begelman 2003).  A jet produced by the accreting BH burrows through the collapsing star, producing a channel through which the relativistic outflow can then escape (MacFadyen $\&$ Woosley 1999; Zhang et al.~2004; Morsony et al.~2007).  In order for a centrifugally-supported disk to form, the core of the stellar progenitor must itself be rapidly rotating; since only a small fraction of all massive stars are likely to satisfy this criterion (e.g., Cantiello et al.~2007; Langer et al.~2008), this provides an explanation for why GRBs are such a rare phenomena.  The duration of the GRB in this model is presumably similar to the $\sim 100$ second in-fall time of the progenitor's He core.

The original collapsar model envisioned a ``failed'' SN, in which most of the star accretes onto the BH instead of becoming unbound.  Somewhat ironically, primary support for the general collapsar picture has come from the association of some long-duration GRBs with bright, ``hyper-''energetic ($E_{\rm SN} \sim 10^{52}$ ergs) Type Ic SNe (e.g. Stanek et al.~2003; Modjaz et al.~2006), which originate from stellar progenitors that have lost their outer H/He envelopes (i.e., Wolf-Rayet stars).  The association of long GRBs with regions of massive star formation (Bloom et al.~1999; Fruchter et al.~2006) has also been key to establishing that most (and possibly all) long-duration GRBs are produced by the core-collapse of massive stars (Woosley $\&$ Bloom 2006).  

Since energetic SNe {\it are} in fact observed coincident with GRBs (i.e. they are hardly a complete ``failure''), if the collapsar picture is correct this implies one or both of the following (cf.~MacFadyen et al.~2001): 
\begin{enumerate}
\item{The explosion mechanism associated with GRB-SNe is fundamentally different than that associated with the death of ``normal'' (slower rotating) massive stars.  MacFadyen $\&$ Woosley (1999), for instance, propose that powerful winds from the accretion disk itself may explode the star.\footnote{A disk wind is probably not the mechanism for most core-collapse SNe, however, since accretion would spin-up the NS and the inferred rotation rates of pulsars at birth are typically much too low (e.g., Kaspi $\&$ Helfand 2002).}  It has been argued, however, that GRB-SNe are not markedly different that non-GRB Type Ic SNe in either their $^{56}$Ni yield or peak brightness (e.g. Soderberg et al.~2006), although GRB-SNe {\it are} more energetic than ``typical'' Type Ic SNe.}
\item{The BH forms only after a delay, due to the ``fallback'' of material that remains gravitationally bound despite a successful SN (e.g. Chevalier 1993).  In this case, the central compact object necessarily initially goes through a NS phase (which must have a decisive role in powering the SN), but the GRB is produced later, following BH formation.  The delay until accretion and collapse cannot be too long in this case, however, (i.e. within $\sim 10-100$ seconds after core bounce) or the accretion rate onto the newly-formed BH would be too low to explain a typical GRB luminosity.}
\end{enumerate}

Although long-duration GRBs are definitively associated with the deaths of massive stars, this does not establish that the CE is a BH.  Single star evolutionary calculations with SN explosions put in by hand suggest that stars with initial masses greater than $\sim 25 M_{\sun}$ form BHs instead of NSs (Woosley $\&$ Weaver 1995).  However, this conclusion depends on the detailed SN mechanism/energy and the stellar progenitor's pre-collapse structure, both of which are poorly understood (see below).  Indeed, our present {\it observational} understanding of the mapping between high mass stars, the Wolf-Rayet progenitors of GRBs, and their compact-object progeny is far from complete (e.g., Smith $\&$ Owocki 2006).  

A satisfactory understanding of the origin of long-duration GRBs ultimately requires an understanding of the SN mechanism(s) as a function of progenitor mass, rotation, metallicity, and binarity.  Modern core collapse simulations find that the outward-propagating shock produced at core bounce initially stalls at a radius $\sim 200$ km due to neutrino and photodissociation losses.  It has long been suspected that neutrino heating from the proto-NS may be crucial to shock ``revival'' and a successful explosion (e.g. Bethe $\&$ Wilson 1985).  However, spherically symmetric simulations fail to produce an explosion (e.g. Mezzacappa et al.~2001) and although more accurate and physical simulations are continually being implemented (e.g. Scheck et al.~2006) and multi-dimensional effects are likely a crucial ingredient (e.g. Murphy and Burrows 2009), it seems unlikely that neutrino heating alone is capable of powering the $\sim 10^{52}$ erg hypernovae associated with long GRBs (e.g. Burrows et al.~2007).  

The massive stellar progenitors of GRBs are, however, far from typical.  The requirements for the collapsar (and, indeed, any GRB model) are rapid rotation and a strong, large-scale magnetic field ($\gtrsim 10^{15}$ G).  The later is required to produce a stable relativistic jet (e.g. McKinney et al.~2006; Tchekovskoy et al.~2008),\footnote{Neutrino annihilation along the rotation axis of the black hole is another possibility for powering a relativistic outflow (e.g. Zaldamea $\&$ Beloborodov 2008).  However, the efficiency of this process is relatively limited, and it may have difficulty explaining very luminous GRBs.} and rapid rotation and strong magnetic fields likely go hand-in-hand in core-collapse: differential rotation provides a source of free energy to power field growth, via e.g. an $\alpha-\Omega$ dynamo in the convective proto-NS (Duncan $\&$ Thompson 1992) or the magneto-rotational instability (MRI; e.g. Akiyama et al.~2003).

\begin{figure} 
    \plottwo{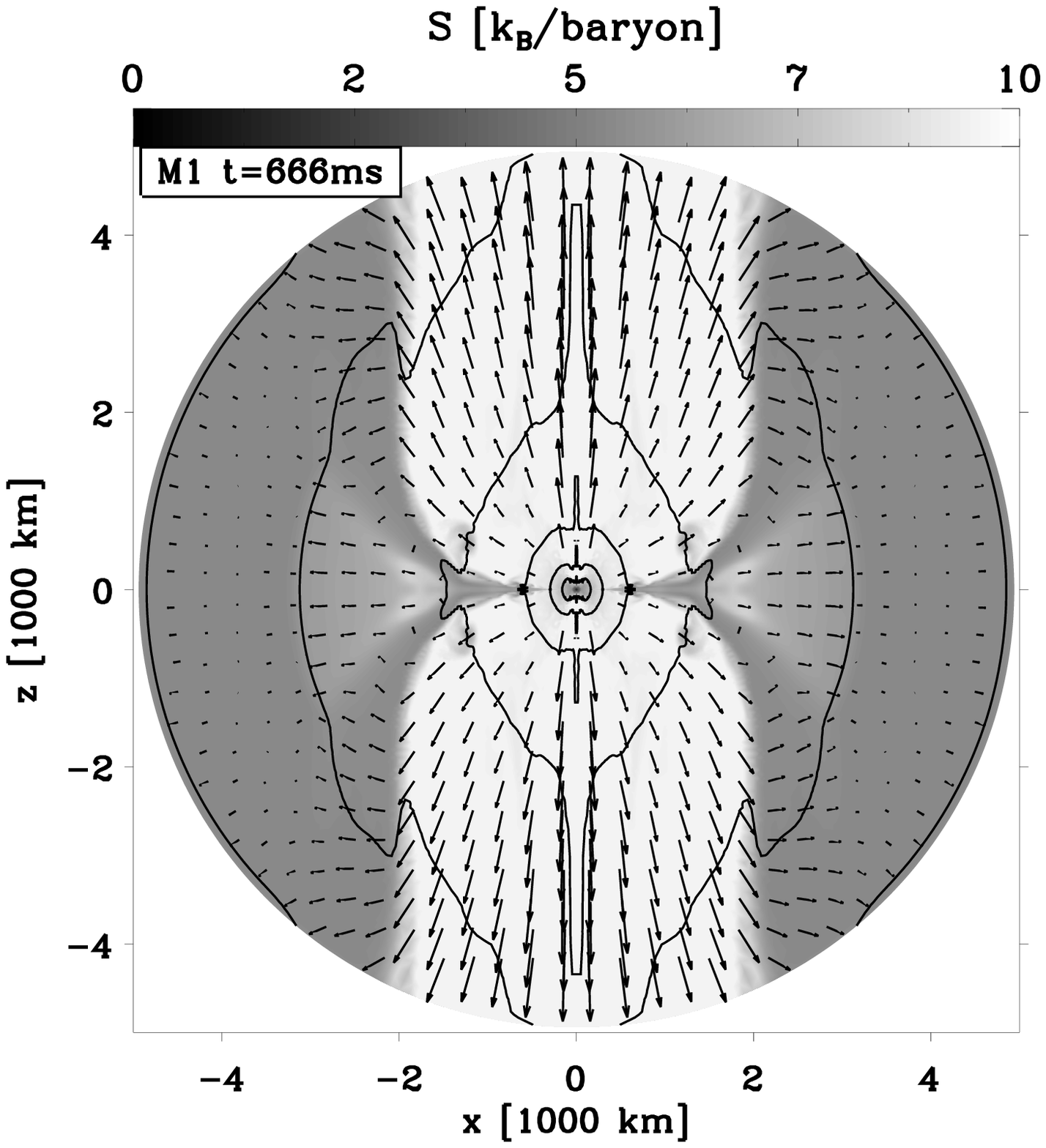}{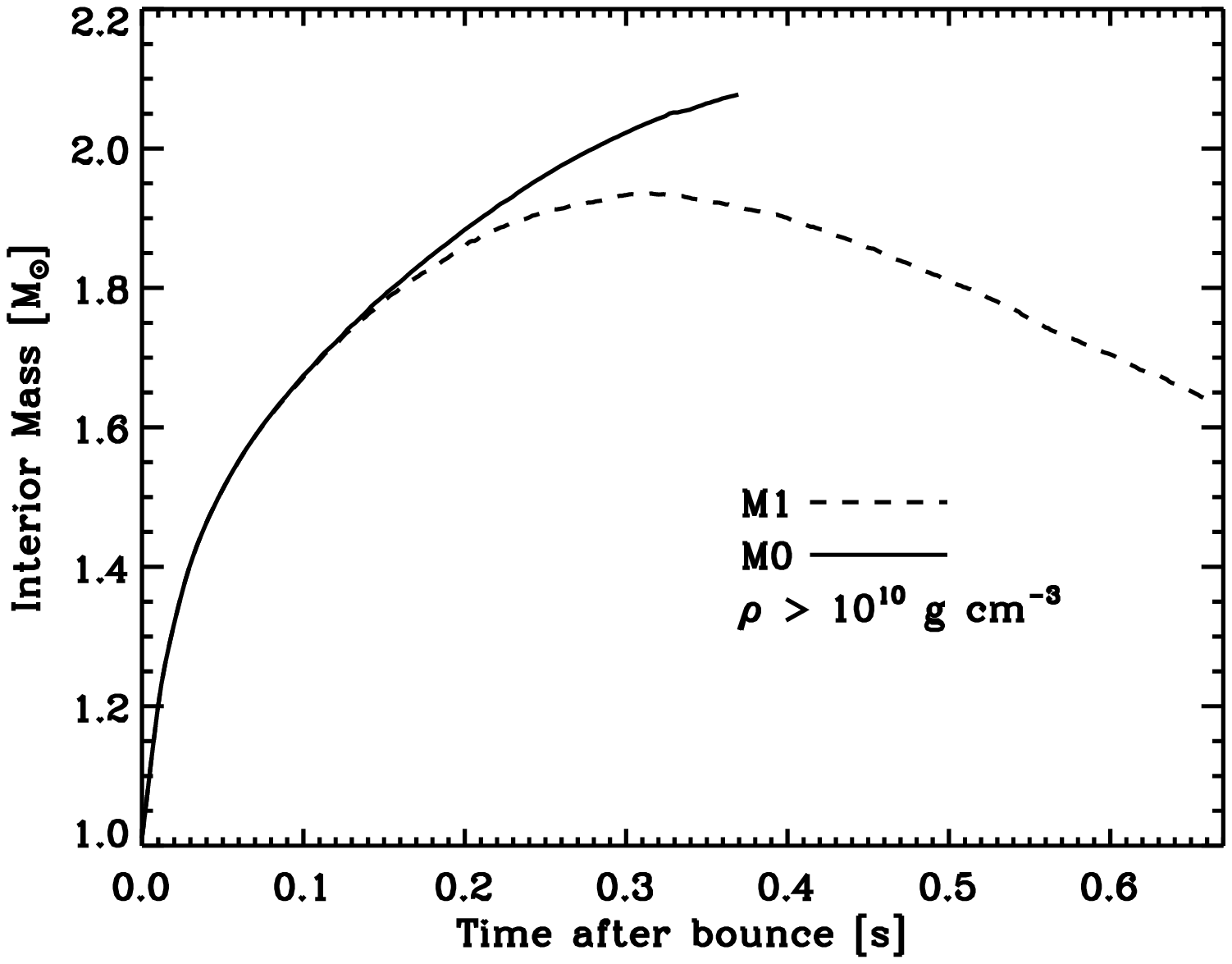}
    \caption{{\it Left Panel:} Entropy map at $t = 666$ ms after core bounce, from the magnetized, rotating ``M1'' core-collapse calculation of Dessart et al.~(2008) using the 35$M_{\sun}$ ZAMS collapsar progenitor model of Woosley $\&$ Heger (2006).  Note the presence of a bipolar MHD outflow, simultaneous with continued accretion of the stellar envelope through a disk in the equator.  {\it Right Panel:} Total proto-NS mass $M_{\rm NS}$ as a function of time after core bounce.  $M_{\rm NS}$ initially grows due to accretion.  However, for model M1 the (magneto-centrifugally-driven) mass loss at high latitudes eventually exceeds the accretion rate and $M_{\rm NS}$ begins to decrease.  BH formation is thus unlikely.  Reproduced with the kind permission of L.~Dessart and {\it The Astrophysical Journal}. }
    \label{fig:Dessart1}
\end{figure}
The crucial question then arises:  {\it \bf Do SNe indeed ``fail'' and lead to BH formation when their progenitor core is rapidly rotating (and, as a likely consequence, possesses a strong magnetic field following collapse)?''} or stated more directly: {\it \bf Are the requisite initial conditions for the collapsar model self-consistent?}

The additional energy reservoir (rotation) and means for extracting it (magnetic fields) make magneto-rotational effects a much more promising avenue for producing energetic ``hypernovae'' than neutrinos alone.  While studies of magnetic fields and rapid rotation in core collapse have a long and storied history (e.g. LeBlanc $\&$ Wilson 1970; Bisnovatyi-Kogan 1970; Symbalisty 1984; Wheeler et al.~2000), only recently have simulations captured both the effects of MHD and neutrino transport (e.g. Burrows et al.~2007; Dessart et al.~2008; hereafter D08).  D08 calculate the collapse of a rotating 35OC progenitor model (35$M_{\sun}$ ZAMS) and collapsar candidate of Woosley $\&$ Heger (2006) using a 2D axisymmetric MHD code with flux-limited neutrino diffusion.  They endow the pre-collapse core with a magnetic field that results in a $\sim 10^{15}$ G field strength when compressed to nuclear densities, thus mimicking the magnetic field expected from the saturated state of the MRI (their simulations do not resolve the MRI itself).  

A sample of D08's results are illustrated in Figure \ref{fig:Dessart1} (cf.~Burrows et al.~2007).  Soon after the collapse to nuclear densities, a bipolar MHD outflow develops from the newly-formed proto-NS.  The explosion is not initially successful over all solid angles, as matter continues to accrete through a disk near the equator.  However, the angular momentum accreted by the NS maintains its rapid rotation, which enhances the neutrino-driven mass-loss rate from mid-latitudes due to magneto-centrifugal ``slinging'' (e.g. Thompson, Chang, $\&$ Quataert 2004; Metzger et al.~2007).  Importantly, in their strongly magnetized model (M1), the mass-loss rate at high latitude eventually exceeds the accretion rate and for $t \gtrsim 300$ ms the NS's mass begins {\it decreasing}!  Although D08's simulations cannot address the possibility of later fall-back, and a different progenitor angular momentum profile could change the conclusions, their results are nonetheless suggestive: a core {\it self-consistently} endowed with the requisite properties to produce a GRB in the collapsar model (strong B and rapid rotation) may not result in a BH at all.\footnote{Note that this conclusion does not imply that BHs cannot form from very massive stars.  It {\it does} suggest that BHs may not form from the subset of rapidly-rotating progenitors which are responsible for hyper-energetic SNe and GRBs.}  

If a BH is not created following core collapse, then a rapidly spinning, highly magnetized proto-NS (a ``proto-magnetar'') likely remains behind in the cavity produced by the (successful) outgoing SN shock.  This naturally leads to a second possibility for producing a long-duration GRB: the ``millisecond magnetar'' model.  

\section{Proto-Magnetar Model}  
\label{sec:magnetar}

Usov (1992) proposed that GRBs may be powered by the spin-down of a newly-formed, rapidly-spinning magnetar.  The baseline motivations for the millisecond magnetar model are two-fold: (1) the rotational energy of a maximally-spinning NS (spin period $P \approx 1$ ms) of radius R is $E_{\rm rot} \sim 3\times 10^{52}(P/1{\,\rm ms})^{-2}(R/10{\,\rm km})^{2}$ ergs, more than sufficient to explain the energetics of most GRBs; and (2) the vacuum/force-free electro-magnetic spin-down power of the NS for a assumed (equatorial) surface dipole field strength $B_{\rm dip}$ is approximately (Spitkovksy 2006)
\begin{equation}
\dot{E}_{\rm FF} \approx \frac{4\pi^{4}B_{\rm dip}^{2}R^{6}}{c^{3}P^{4}} \approx 10^{49}{\,\rm ergs\,s^{-1}}\left(\frac{B_{\rm dip}}{10^{15}{\rm G}}\right)^{2}\left(\frac{P}{1{\,\rm ms}}\right)^{-4}\left(\frac{R}{10{\,\rm km}}\right)^{6},
\label{eq:edotFF}
\end{equation}
similar to observed (beaming-corrected) GRB luminosities $\dot{E} \sim 10^{49}-10^{51}$ ergs s$^{-1}$ for $P \sim 1$ ms and $B_{\rm dip} \sim 10^{15}-10^{16}$ G (Blackman $\&$ Yi 1998; Wheeler et al.~2000).  Though initially speculative because observed radio pulsars generally have $B_{\rm dip} \lesssim 10^{14}$ G (Manchester 2004), evidence for the existence of magnetars has increased significantly in the past decade (Kouveliotou et al.~1998; Hurley et al.~2005).  Indeed, Woods $\&$ Thompson (2006) argue that $\sim 10\%$ of Galactic NSs are formed with $B_{\rm dip} \gtrsim 10^{14}$ G, and the location of some magnetars within massive star clusters suggests that they may originate from massive stellar progenitors ($\gtrsim 40M_{\sun}$; Muno et al.~2006). 

Magnetars are not, however, born in vacuum, nor are their magnetospheres necessarily well-described as force-free at early times.  Like other NSs, magnetars are born hot and emit a substantial neutrino luminosity as they radiate their gravitational binding energy $\sim 3\times 10^{53}$ ergs (e.g. Burrows $\&$ Lattimer 1987).  Despite its relatively weak interaction, this huge neutrino flux drives baryons from the surface of the magnetar (primarily via $\nu_{e}/\bar{\nu}_{e}$ absorption) during the first $\sim 10-100$ seconds of its life (Thompson, Chang, $\&$ Quataert 2004; Metzger et al.~2007).  Such a ``neutrino-heated wind'' is a generic consequence of a successful SN (e.g. Burrows et al.~1995).  Indeed, Qian $\&$ Woosley (1996) show that the neutrino-driven mass-loss rate $\dot{M}_{\nu}$ from a non-magnetized, non-rotating proto-NS is approximately given by  
\begin{equation}
\dot{M}_{\nu} = 10^{-4}M_{\sun}{\,\rm s^{-1}}\left(\frac{L_{\nu}}{10^{52}{\,\rm ergs\,s^{-1}}}\right)^{5/3}\left(\frac{T_{\nu}}{3{\,\rm MeV}}\right)^{10/3}\left(\frac{M}{1.4M_{\sun}}\right)^{-2}\left(\frac{R}{10{\,\rm km}}\right)^{5/3},
\label{eq:mdotQW}
\end{equation}
where $M$, $R$, $L_{\nu}$, and $T_{\nu}$ are the proto-NS's mass, radius, electron neutrino luminosity, and neutrinosphere temperature, respectively.

Metzger et al.~(2007; hereafter M07) study the effects of magnetic fields and rotation on proto-NS winds by solving the equations of neutrino-heated MHD in the equatorial plane of the proto-NS (somewhat analogous to the Weber $\&$ Davis 1967 model of the solar wind).  By encorporating the cooling evolution of the proto-NS (e.g. $L_{\nu}$(t) and $T_{\nu}$(t) from Pons et al.~1999) and calibrating their results using multi-dimensional MHD simulations (Bucciantini et al.~2006), M07 calculate how the mass-loss rate $\dot{M}$ and energy-loss rate $\dot{E}$ of magnetars evolve during the first $\sim 100$ seconds following their formation.  From $\dot{M}(t)$ and $\dot{E}(t)$, they determine the ``magnetization'' $\sigma(t) \equiv \dot{E}/\dot{M}c^{2}$.  For ultra-relativistic outflows ($\sigma \gg 1$), $\sigma$ represents the maximum potential Lorentz factor $\Gamma_{\rm max}$, obtained if the outflow's Poyting flux is completely converted into kinetic power.  Thus, a GRB-producing outflow must have $\sigma \gtrsim 10^{2}-10^{3}$ to overcome compactness constraints ($\S\ref{sec:obs}$).  

\begin{figure}
\plottwo{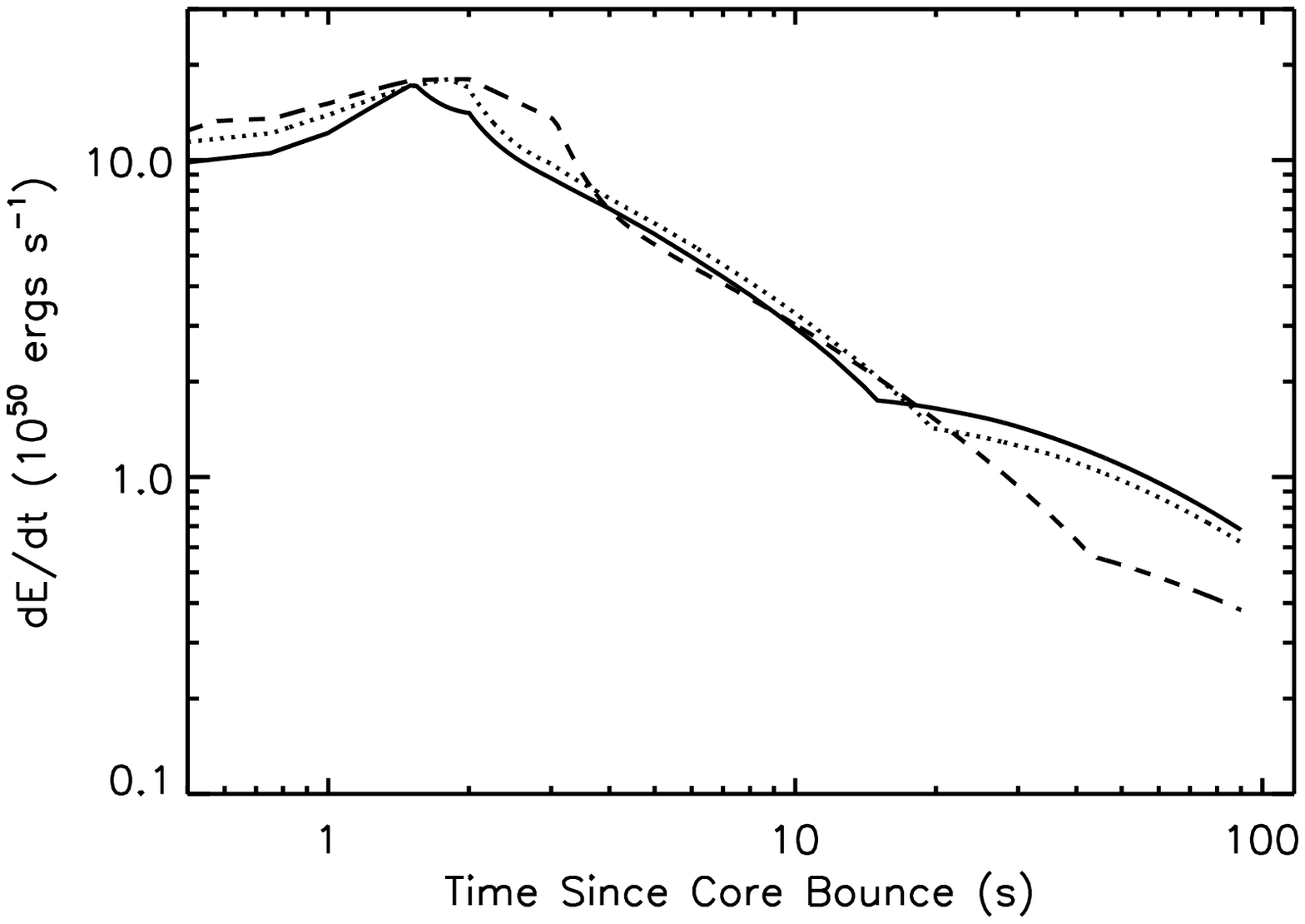}{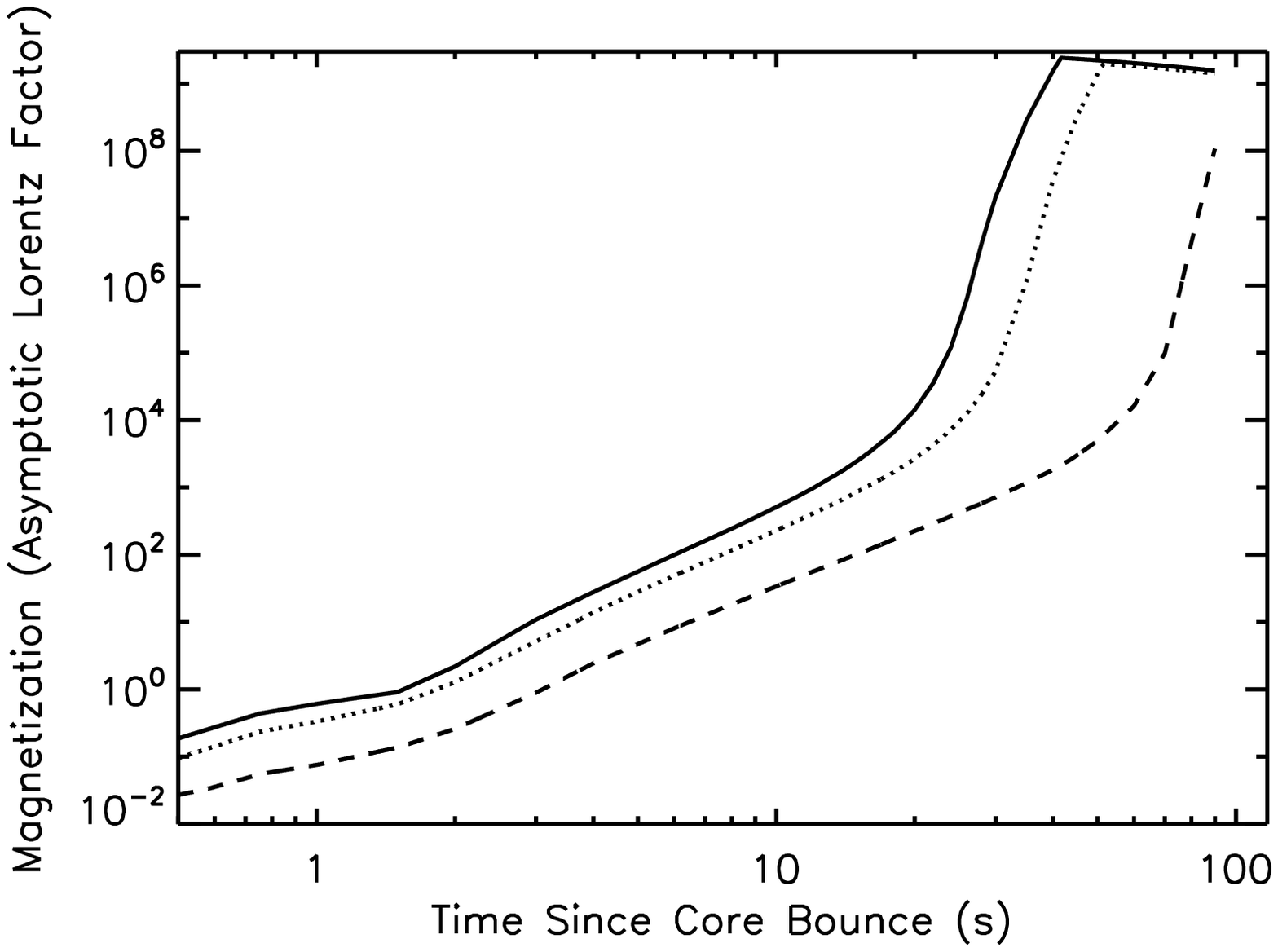}
\caption{Energy loss rate $\dot{E}$ and magnetization ($\sigma = \Gamma_{\rm max} \equiv \dot{E}/\dot{M}c^{2}$) as a function of time since core bounce, calculated for a proto-magnetar with initial spin period $P = 1$ ms, fixed surface dipole field strength $B_{\rm dip} = 3\times 10^{15}$ G, and three values of the NS mass $M_{\rm NS} = 1.2M_{\sun}$ ({\it dashed line}), $1.4M_{\sun}$ ({\it dotted line}), and $2.0M_{\sun} ({\it solid line})$ (from the model of Metzger et al.~2007).  This Figure illustrates that millisecond proto-magnetars produces outflows with the correct power ($\dot{E} \sim 10^{50}$ ergs s$^{-1}$) and magnetization (maximum Lorentz factor $\Gamma_{\rm max} \sim 10^{2}-10^{3}$) on the correct timescale ($t\sim 10-100$ seconds) to explain long-duration GRBs.  The abrupt rise in $\sigma$ at $t \sim 30-100$ second, which occurs when the proto-NS becomes optically-thin to neutrinos, provides a possible explanation for the observed sharp cut-off in GRB prompt emission (e.g. Barniol-Duran $\&$ Kumar 2009).}
\label{fig:metzger07}
\end{figure}

\begin{figure}
\plottwo{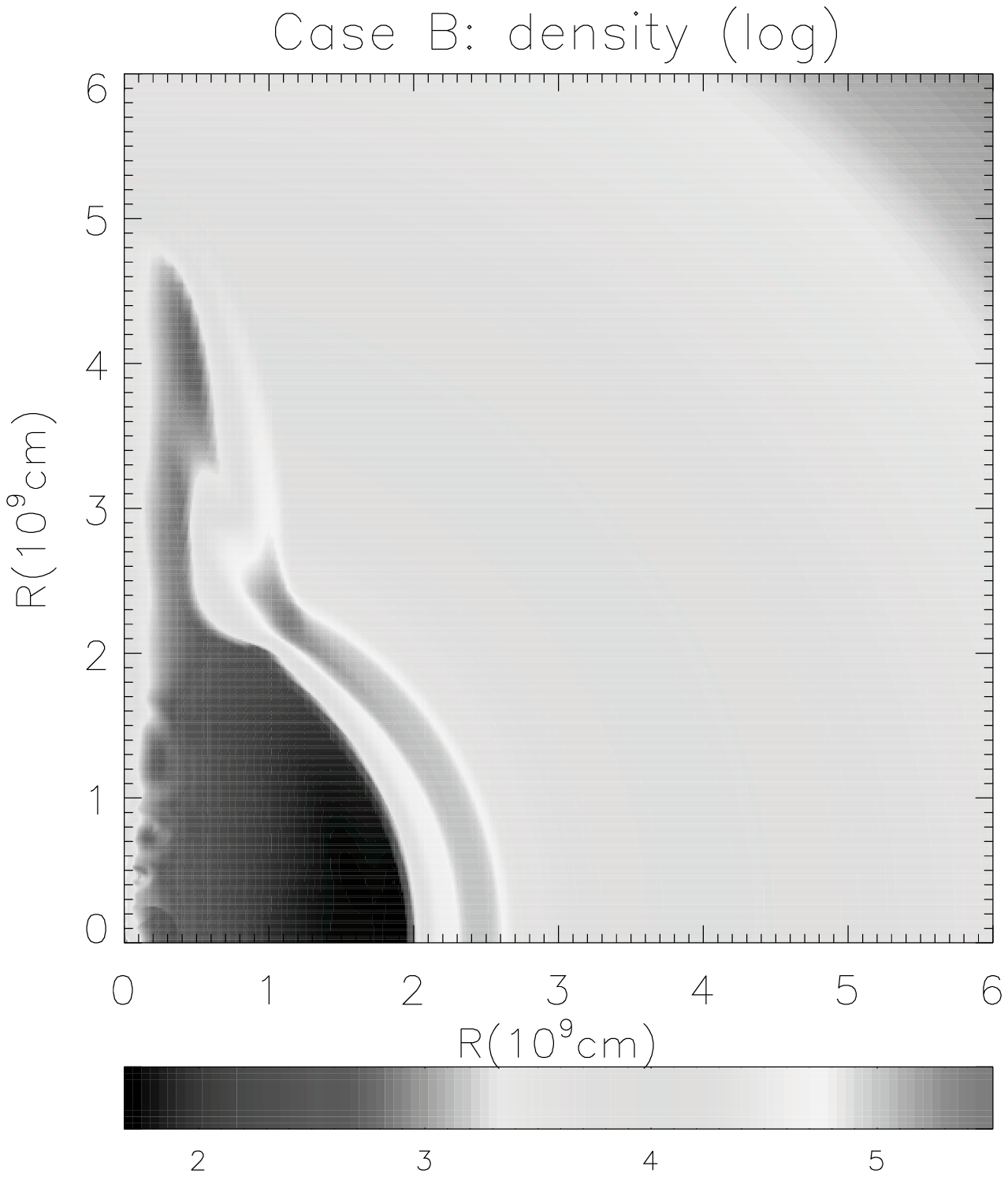}{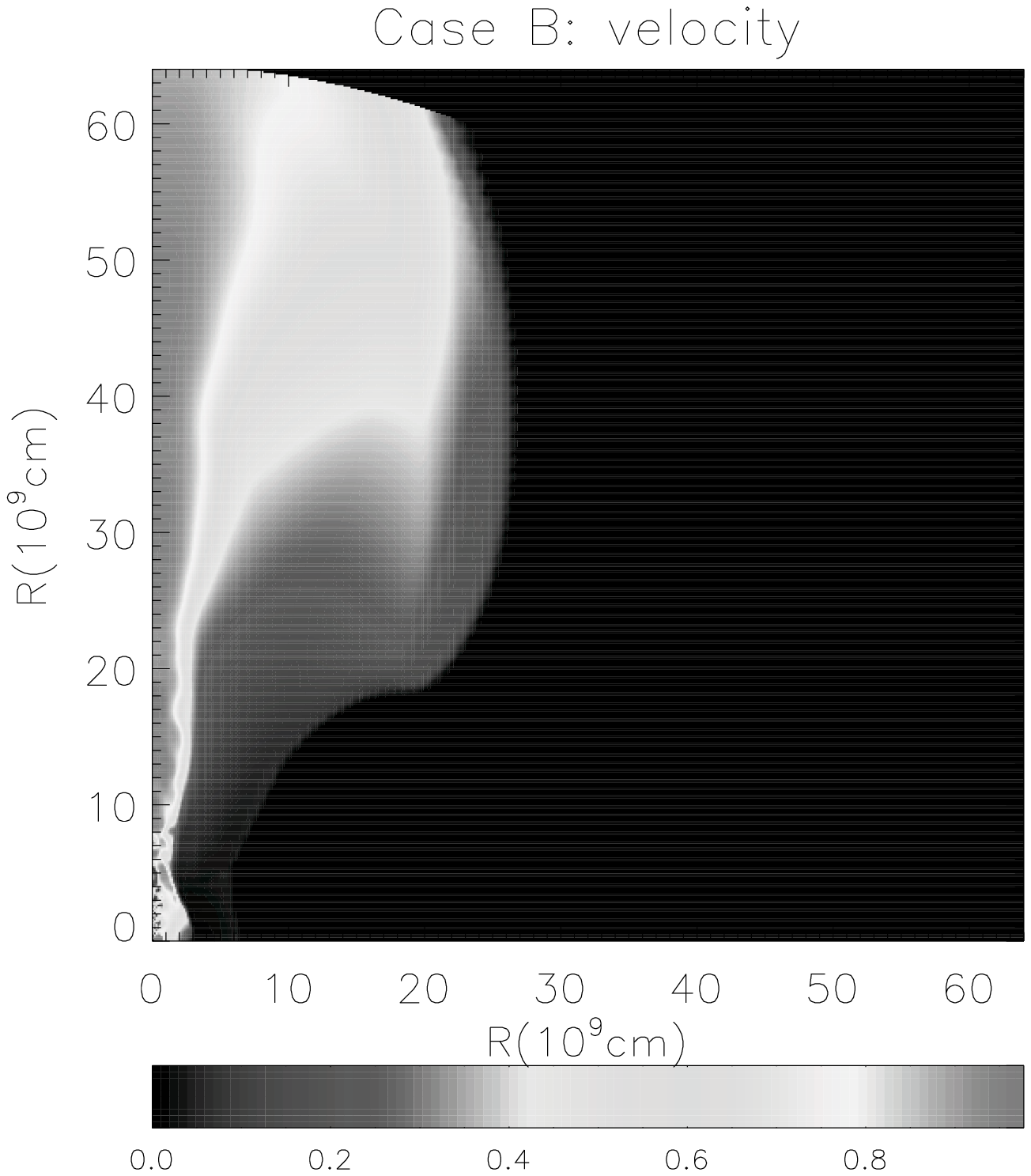}
\caption{Evolution of the wind from a $B_{\rm dip} = 3\times 10^{15}$ G, $P = 1$ ms proto-magnetar, released inside an exploding 35$M_{\sun}$ ZAMS star, from the axisymmetric MHD calculations of Bucciantini et al.~(2009).  The wind is injected at small radii with a time-dependent power $\dot{E}(t)$ and magnetization $\sigma(t)$ from the calculations of Metzger et al.~(2007; see Fig.~\ref{fig:metzger07}).  {\it Left Panel:} Density profile [log(g cm$^{-3}$)] at $t = 2$ seconds after core bounce.  A magnetized nebula develops behind the SN shock, whose anisotropic pressure begins carving a bipolar channel along the rotation axis.  {\it Right Panel:} Velocity at $t = 5$ seconds in units of $c$ ({\it zoomed out}). A complete channel opens along the polar axis.  The nebular magnetic field redirects the proto-magnetar's outflow into an ultra-relativistic, bipolar jet with an opening angle $\theta_{\rm jet} \sim 5-10^{\circ}$.}
\label{fig:B09}
\end{figure}

Figure \ref{fig:metzger07} shows calculations of $\dot{E}(t)$ and $\sigma(t)$ from an updated version of M07's model for a proto-magnetar with $B_{\rm dip} = 3\times 10^{15}$ and an initial spin period $P = 1$ ms, calculated for three different values of the NS mass ($M = 1.2, 1.4$, and 2.0 $M_{\sun}$).  Initially $\dot{E}$ rises as the proto-NS contracts and spins-up, but $\dot{E}$ secularly decreases for $t \gtrsim  2$ s once the NS begins to spin down.  At early times ($ t \sim $ few seconds) the proto-magnetar wind goes through a powerful ($\dot{E} \sim 10^{52}$ ergs s$^{-1}$) non-relativistic ($\sigma < 1$) phase, which plays a crucial role in carving a bipolar channel through the stellar progenitor's envelope (see Fig.~\ref{fig:B09}), but is too mass-loaded to produce a GRB.  At later times, however, $\sigma \propto \dot{M}^{-1}$ rises rapidly because $\dot{M} \propto L_{\nu}^{5/3}T_{\nu}^{10/3}$ decreases quickly as the NS cools.  The proto-magnetar wind becomes ultra-relativistic ($\sigma = \Gamma_{\rm max} \gtrsim 100-1000$) on a timescale $t \sim 10-100$ seconds, comparable to the duration of a typical GRB. 

We illustrate the robustness of this important result by estimating $\sigma$ explicitly.  Although at early times magneto-centrifugal enhancements to $\dot{M}_{\nu}$ (eq.~[\ref{eq:mdotQW}]) can be significant and $\dot{E}_{\rm FF}$ (eq.~[\ref{eq:edotFF}]) is not accurate for $\sigma \lesssim 1$, at later times (once the NS spins down a bit and $\sigma \gg 1$), equations (\ref{eq:mdotQW}) and (\ref{eq:edotFF}) are reasonable approximations.  The magnetization when $\sigma \gg 1$ is therefore 
\begin{eqnarray} 
\sigma|_{\sigma \gg 1} = \Gamma_{\rm max} \approx \left(\frac{\dot{E}_{\rm FF}}{\dot{M}_{\nu}c^{2}}\right)\times\left(\frac{R_{\rm NS}}{2R_{\rm L}}\right)^{-1} \nonumber 
\end{eqnarray}
\begin{eqnarray}
\approx 10^{3}\left(\frac{B_{\rm dip}}{10^{15}{\rm G}}\right)^{2}\left(\frac{P}{2{\,\rm ms}}\right)^{-3}\left(\frac{L_{\nu}}{10^{50}{\,\rm ergs\,s^{-1}}}\right)^{-5/3}\left(\frac{T_{\nu}}{2{\,\rm MeV}}\right)^{-10/3}\left(\frac{M}{1.4M_{\sun}}\right)^{2}\left(\frac{R}{10{\,\rm km}}\right)^{10/3}
\label{eq:sigma}
\end{eqnarray}
where $R_{\rm L} = 2\pi Pc$ is the light cylinder radius, the factor $R_{\rm N}/2R_{\rm L}$ is the fraction of the NS surface threaded by open magnetic flux for an aligned rotator (i.e. we assume only the ``open zone'' of the magnetosphere contributes to $\dot{M}_{\nu}$), and we have scaled $L_{\nu}$ and $T_{\nu}$ to typical values at $t \approx 30$ seconds after core-bounce (Pons et al.~1999).  Equation (\ref{eq:sigma}) shows that for the same magnetic fields and rotation rates necessary to power long GRBs ($B_{\rm dip} \gtrsim 10^{15}$ G, P $\sim$ 1 ms), the magnetar model naturally predicts the correct Lorentz factors ($\Gamma \sim 10^{2}-10^{3}$).  Equation (\ref{eq:sigma}) is robust because $\dot{M}_{\nu}$ (eq.~[\ref{eq:sigma}]) results from well-understood {\it thermal} physics and weak interactions at the proto-NS surface (Qian $\&$ Woosley 1996; eq.~[\ref{eq:mdotQW}]) and the basic electro-magnetic spin-down of a force-free rotator is now a solved theoretical problem (e.g. Spitkovksy 2006; eq.~[\ref{eq:edotFF}]).

At later times ($t \gtrsim 30-100$ seconds) the proto-NS becomes optically thin to neutrinos and, consequentially, the neutrino-driven mass loss rate effectively vanishes.  As this occurs $\sigma \propto \dot{M}^{-1}$ rapidly rises until finally plateauing at a very high value ($\sigma \gtrsim 10^{9}$), once the mass-loss is dominated by electron/positron pairs driven by strong electric fields (e.g. Arons $\&$ Scharlemann 1979) rather than neutrino heating.  Since radiative processes likely become inefficient for ultra-high-$\sigma$ flows (e.g. internal shocks move to very large radii and magnetic dissipation becomes slow; e.g. Giannios $\&$ Spruit 2005), this $\nu$-optically-thin transition plausibly signals the end of the prompt GRB phase in the magnetar model.  The abrupt rise in $\sigma$ at $t \sim 30-100$ s may thus explain why prompt GRB emission is observed to cut off abruptly (e.g. Barniol-Duran $\&$ Kumar 2009).  

Although the above discussion illustrates that proto-magnetars produce outflows with the correct characteristics to explain long-duration GRB, it does not address how the outflow escapes from the surrounding, exploding star and becomes collimated into a bipolar jet.  Using axisymmetric MHD calculations, Bucciantini et al.~(2007, 2008, 2009) suggest a model for the collimation of proto-magnetar outflows which is similar to that applied to explain the evolution and morphology of (much older) pulsar wind nebulae such as the Crab Nebula (e.g., Begelman $\&$ Li 1992).  The essential point is that the (exploding) star {\it inertially confines} the proto-magnetar wind (cf.~Uzdensky $\&$ MacFadyen 2006; Komissarov $\&$ Barkov 2007), causing toroidal magnetic flux to accumulate in a shocked nebula behind the outgoing SN shock.  Magnetic hoop stresses confine the gas pressure in this ``proto-magnetar nebula'' along the polar axis.  This in turn exerts an anisotropic pressure on the SN shell (the so-called ``tube of toothpaste'' effect), which rapidly opens a polar cavity that directs and collimates the proto-magnetar wind into a bipolar jet.  Snapshots of this process from Bucciantini et al.~(2009) are shown in Figure \ref{fig:B09}.  This represents the first self-consistent GRB jet calculation that extends from the central engine to the stellar surface and which is evolved for a time $\Delta t \sim 10$ s comparable to the duration of a GRB.  Importantly, these calculations illustrate that once the jet breaks out of the star, its mass and energy flux map nearly one-to-one onto those injected by the proto-magnetar wind at small radii.  This justifies using $\dot{E}(t)$ and $\sigma(t)$ calculated using a free wind model (such as those in Fig.~\ref{fig:metzger07}) as direct input into a GRB emission model.

\begin{table}
%\begin{scriptsize}
\begin{center}
\vspace{0.05 in}\caption{Black Hole vs. Proto-Magnetar Long-Duration GRB Models}
\label{table:comp}
\begin{tabular}{lccc}
\hline
\hline
\multicolumn{1}{c}{Property} &
\multicolumn{1}{c}{Collapsar (Black Hole)} &
\multicolumn{1}{c}{Proto-Magnetar} &
\\
\hline
Total Energy &  $E_{\rm max} \sim 0.1M_{\star}c^{2}$ & $E_{\rm max} \sim 3\times 10^{52}$ ergs \\
 & $\sim 10^{54}$ ergs (Huge) & (Max NS Rotation Rate)  \\
\hline
Duration & Stellar Envelope & Spin-Down Time OR \\
 & Infall Time & Until Proto-NS $\nu$-Optically Thin \\
\hline 
Lorentz Factor & Neutron Diffusion? & $\Gamma_{\rm max} \sim 10^{2}-10^{3}$ $@$ $t \sim 10-100$ s \\
 & {\scriptsize (Levinson $\&$ Eichler 1999)} & Higher L$_{\gamma} \leftrightarrow$ Higher $\Gamma_{\rm max}$ \\
\hline
Time-Averaged & Tracks Accretion & Single Envelope (Fig.~\ref{fig:lc}) if: \\
 Light Curve Shape& Rate? & {\scriptsize (1) internal shocks}  \\
 & & {\scriptsize (2) minimal jet-envelope interaction} \\
\hline
Collimation & {\bf Accretion Disks} & Jet Formation via Stellar \\
 & {\bf Produce Jets} & Confinement (see Fig.~\ref{fig:B09}) \\ 
\hline
Supernova & Accretion Disk Winds & {\bf Neutron Stars} \\
Association & or Delayed BH Formation & {\bf Power Supernovae} \\
\hline
Late-Time & Late-Time Fall-Back & {\bf Magnetar Remains} \\
X-Ray Flaring & Accretion & & \\
\hline
Particle  & e$^{-}/e^{+}$ Pairs  & Baryons for $t \lesssim 100$ s \\
Composition & or Baryons (?) &  Pairs for $t \gtrsim 100$ s \\
\hline
\end{tabular}
\end{center}
\end{table}

\begin{figure}
\begin{center}
\scalebox{0.4}{\includegraphics{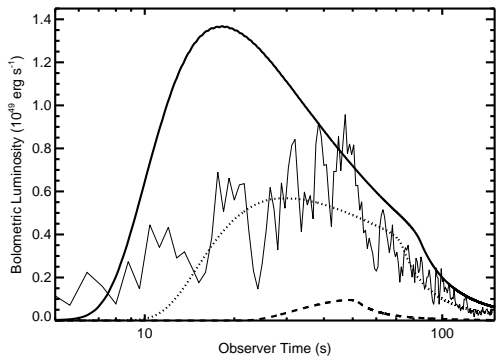}}
%\plotone{fig2.eps}
\caption{Luminosity of internal shock emission in proto-magnetar winds for a NS with initial rotation period $P = 1$ ms and three surface dipole magnetic field strengths: $B_{\rm dip} = 10^{15}$ G (\emph{dashed line}), $3\times 10^{15}$ G (\emph{dotted line}), and $10^{16}$ G (\emph{solid line}).  The gradual onset of the emission begins once the magnetization of the outflow $\sigma$ reaches $\gtrsim 10$ due to the large Thomson optical depth, which decreases as the outflow expands.  The light curve was calculated for ``isolated'' magnetar birth (e.g. a quasi-isotropic outflow).  Correcting for the collimation due to the surrounding stellar envelope (Fig.~\ref{fig:B09}), the {\it observed} luminosity is larger by a factor of $\sim \theta_{j}^{-2} \sim 10^{2}$.  The late-time BAT light curve from GRB 060614 (Gehrels et al.~2006), shown with a light solid line, is reproduced in a time-averaged sense by the $B_{\rm dip} = 3\times 10^{15}$ G model.  From Metzger, Quataert, $\&$ Thompson (2008).}
\label{fig:lc}
\end{center}
\end{figure}

\section{Conclusions and Future Directions}
\label{sec:conclusions}

Observationally distinguishing between BH and magnetar models for long-duration GRBs is a difficult task.  A comparison between the models is summarized in Table \ref{table:comp}.  In principle, a detailed comparison could be performed between theoretical models and the observed prompt emission.  Figure \ref{fig:lc} shows a prediction for the light curve expected from ``isolated'' magnetar birth (i.e. neglecting the effects of an overlying stellar envelope), assuming that radiation is generated as the free energy in the outflow is dissipated through internal shocks.  The qualitative shape of the light curve resembles the time-averaged ``envelope'' of a surprisingly large fraction of long GRBs.  Unfortunately, making definitive predictions, even when the initial properties of the jet are known, is hindered by our ignorance of the dissipation and radiation mechanisms in the outflow (recall the unanswered questions from $\S1$).  

The magnetar model does, however, make a few robust (albeit indirect) predictions.  For instance, (1) at $t \sim 10$ seconds after core-collapse, {\bf outflows from higher luminosity bursts should possess higher Lorentz factors}: although the spin-down luminosity is larger for more rapidly spinning, highly magnetized NSs, the mass-loss rate at this epoch is probably similar for all NSs (see eqs.~$[\ref{eq:mdotQW}]-[\ref{eq:sigma}]$).  Furthermore, (2) {\bf the Lorentz factor should be a monotonically-increasing function of time during the burst} (Fig.~\ref{fig:metzger07}).  The magnetar model also predicts that (3) {\bf for the first $\sim 10-100$ s after core bounce (when the neutrino luminosity is large), the GRB-producing outflow should be baryon-dominated}, unlike the pair-dominated outflows expected from radio pulsars (Arons $\&$ Scharlemann 1979) or along flow-lines that thread the event horizon of a BH.  If a baryon-rich outflow can be definitively ruled out on this early timescale (e.g. via searches for neutrino emission due to hadronic processes with, e.g., IceCube; Abbasi et al.2009), the magnetar model could be refuted.  

Unfortunately, similar predictions are not yet available for BH models, in part because it is still not clear how the mass and energy fluxes in jets from accretion disks are determined.  Indeed, the jet's mass-loading may depend on uncertain diffusive processes from the jet walls (e.g. Levinson $\&$ Eichler 2000; McKinney 2005).  Simply put, more robust predictions are possible from the proto-magnetar model because {\it spheres are simpler than disks} and {\it the surface of a proto-NS is in kinetic equilibrium.}  Given these virtues, one promising strategy to break the present BH-magnetar stalemate is to transform the predictions for $\dot{E}(t)$ and $\sigma(t)$ from the magnetar model (Fig.~\ref{fig:metzger07}) into observable gamma-ray light curves and spectra (e.g. Fig.~\ref{fig:lc}).  Hopefully, then, robust and falsifiable predictions will emerge, and we may be able to discern the central engine by the process of elimination.  This course of action signals a renewed theoretical and observational challenge to determine the source and mechanism of prompt GRB emission. 

\acknowledgements
I thank the organizers and participants of the BASH 2009 Symposium for an interesting and diverse meeting.  Much of the opinions expressed herein were forged as the result of fruitful collaboration and many helpful discussions with a number of individuals, especially E.~Quataert, T.~Thompson, N.~Bucciantini, T.~Piro, J.~Arons, D.~Giannios, D.~Perley, and J.~Bloom.  I thank L.~Dessart for providing Figure $\ref{fig:Dessart1}$ and J.~Pons for sharing his proto-NS cooling calculations.  I acknowledge support by NASA through the Einstein Fellowship Program, grant \#NAS8-03060.

%%% THE BIBLIOGRAPHY
%%%
%%% CONSULT SECTION 3 OF "INSTRUCTIONS FOR AUTHORS" FOR HOW TO USE NATBIB.
%%% AUTHORS ARE ENCOURAGED TO USE EITHER THE "THEBIBLIOGRAPY" ENVIRONMENT
%%% BY UNCOMMENTING (DELETING THE "%" SYMBOL) THE COMMANDS BELOW, OR BY
%%% USING THE BIBTEX ENVIRONMENT. TO FIND OUT WHICH IS APPLICABLE TO YOUR
%%% CONTRIBUTION, CONSULT THE VOLUME EDITORS FOR YOUR PROCEEDINGS.
%%%


\begin{thebibliography}

%\bibitem{1976HEYpJ...210..498B}
\bibitem[]{} IceCube Collaboration: R.~Abbasi 2009, arXiv:0902.0131 

\bibitem[Abdo et al.(2009)]{2009Sci...323.1688A} Abdo, A.~A., et al.\ 2009a, 
Science, 323, 1688 

\bibitem[Abdo et al.(2009)]{2009ApJ...706L.138A} Abdo, A.~A., et al.\ 2009b, 
\apjl, 706, L138 

\bibitem[Akiyama et al.(2003)]{2003ApJ...584..954A} Akiyama, S., Wheeler, 
J.~C., Meier, D.~L., \& Lichtenstadt, I.\ 2003, {\rm ApJ\/}, 584, 954

\bibitem[]{} Arons, J., \& Scharlemann, E.~T.\ 1979, \apj, 231, 854

\bibitem[Barniol Duran 
\& Kumar(2009)]{2009MNRAS.395..955B} Barniol Duran, R., \& Kumar, P.\ 2009, \mnras, 395, 955 

\bibitem[]{} Begelman, M.~C., \& Li, Z.-Y.\ 1992, \apj, 397, 187 

\bibitem[]{} Berger, E., et al.\ 2003, \nat, 426, 154 

\bibitem[Berger(2009)]{2009ApJ...690..231B} Berger, E.\ 2009, \apj, 690, 
231 

\bibitem[]{} Bethe, H.~A., \& Wilson, J.~R.\ 1985, \apj, 295, 14 

\bibitem[Bisnovatyi-Kogan(1970)]{1970AZh....47..813B} Bisnovatyi-Kogan, 
G.~S.\ 1970, \azh, 47, 813 

\bibitem[]{} Blackman, E.~G., \& Yi, 
I.\ 1998, \apjl, 498, L31 

\bibitem[]{} Blandford, R.~D.\ 2002, 
Lighthouses of the Universe: The Most Luminous Celestial Objects and Their 
Use for Cosmology, 381 

\bibitem[]{} Bloom, J.~S., et al.\ 1999, \apjl, 518, L1 

\bibitem[]{} Bucciantini, N., 
Thompson, T.~A., Arons, J., Quataert, E., \& Del Zanna, L.\ 2006, \mnras, 
368, 1717 

\bibitem[]{} Bucciantini, N., Quataert, E., Arons, J., Metzger, B.~D., 
\& Thompson, T.~A.\ 2007, \mnras, 380, 1541 

\bibitem[]{} Bucciantini, N., 
Quataert, E., Arons, J., Metzger, B.~D., 
\& Thompson, T.~A.\ 2008, \mnras, 383, L25 

\bibitem[]{} Bucciantini, N., 
Quataert, E., Metzger, B.~D., Thompson, T.~A., Arons, J., 
\& Del Zanna, L.\ 2009, arXiv:0901.3801 

\bibitem[Burrows et al.(2007)]{2007ApJ...664..416B} Burrows, A., Dessart, 
L., Livne, E., Ott, C.~D., \& Murphy, J.\ 2007, \apj, 664, 416 

\bibitem[]{} Burrows, A., Hayes, J., 
\& Fryxell, B.A.\ 1995, \apj, 450, 830

\bibitem[]{} Burrows, A., \& Lattimer, J.~M.\ 1986, \apj, 307, 178 

\bibitem[]{} Burrows, D.~N., et al.\ 2005, Science, 309, 1833 

\bibitem[Cantiello et 
al.(2007)]{2007A&A...465L..29C} Cantiello, M., Yoon, S.-C., Langer, N., \& Livio, M.\ 2007, \aap, 465, L29 

\bibitem[Cenko et al.(2009)]{2009arXiv0905.0690C} Cenko, S.~B., et al.\ 
2009, arXiv:0905.0690 

\bibitem[Chevalier(1993)]{1993ApJ...411L..33C} Chevalier, R.~A.\ 1993, 
\apjl, 411, L33 

\bibitem[]{} Dessart, L., Burrows, 
A., Livne, E., \& Ott, C.~D.\ 2008, \apjl, 673, L43 

\bibitem[]{} Duncan, R.~C., \& Thompson, C.\ 1992, \apjl, 392, L9 

\bibitem[]{} Fishman, G.~J., \& Meegan, C.~A.\ 1995, \araa, 33, 415 

\bibitem[]{} Frail, D.~A., Kulkarni, 
S.~R., Nicastro, L., Feroci, M., \& Taylor, G.~B.\ 1997, \nat, 389, 261 

\bibitem[]{} Frail, D.~A., et al.\ 2001, \apjl, 562, L55 

\bibitem[]{} Fruchter, A.~S., et al.\ 2006, \nat, 441, 463 

\bibitem[]{} Gehrels, N., et al.\ 2006, \nat, 444, 1044 

\bibitem[Gehrels et 
al.(2009)]{2009ARA&A..47..567G} Gehrels, N., Ramirez-Ruiz, E., \& Fox, D.~B.\ 2009, \araa, 47, 567 

\bibitem[]{} Giannios, D., \& Spruit, H.~C.\ 2005, \aap, 430, 1 

\bibitem[Hurley et al.(2005)]{2005Natur.434.1098H} Hurley, K., et al.\ 
2005, \nat, 434, 1098 

\bibitem[]{} Kaspi, V.~M., \& Helfand, D.~J.\ 2002, Neutron Stars in Supernova Remnants, 271, 3 

\bibitem[Katz(1997)]{1997ApJ...490..633K} Katz, J.~I.\ 1997, \apj, 490, 633 

\bibitem[]{} Klebesadel, R.~W., 
Strong, I.~B., \& Olson, R.~A.\ 1973, \apjl, 182, L85 

\bibitem[Kobayashi et al.(1997)]{1997ApJ...490...92K} Kobayashi, S., Piran, 
T., \& Sari, R.\ 1997, \apj, 490, 92 

\bibitem[Komissarov 
\& Barkov(2007)]{2007MNRAS.382.1029K} Komissarov, S.~S., \& Barkov, M.~V.\ 2007, \mnras, 382, 1029 

\bibitem[]{} Kouveliotou, C., 
Meegan, C.~A., Fishman, G.~J., Bhat, N.~P., Briggs, M.~S., Koshut, T.~M., 
Paciesas, W.~S., \& Pendleton, G.~N.\ 1993, \apjl, 413, L101 

\bibitem[]{} Kouveliotou, C., et 
al.\ 1998, \nat, 393, 235 

\bibitem[Kumar 
\& Barniol Duran(2009)]{2009MNRAS.400L..75K} Kumar, P., \& Barniol Duran, R.\ 2009, \mnras, 400, L75 

\bibitem[Kumar et al.(2008)]{2008MNRAS.388.1729K} Kumar, P., Narayan, R., 
\& Johnson, J.~L.\ 2008, \mnras, 388, 1729 

\bibitem[]{} Langer, N., Cantiello, M., Yoon, S.-C., Hunter, I., Brott, I., Lennon, D., de Mink, S.,\& Verheijdt, M.\ 2008, IAU Symposium, 250, 167

\bibitem[Lazzati 
\& Perna(2007)]{2007MNRAS.375L..46L} Lazzati, D., \& Perna, R.\ 2007, \mnras, 375, L46 

\bibitem[LeBlanc 
\& Wilson(1970)]{1970ApJ...161..541L} LeBlanc, J.~M., \& Wilson, J.~R.\ 1970, \apj, 161, 541 

\bibitem[]{}  Levinson, A., \& 
Eichler, D.\ 2003, \apjl, 594, L19

\bibitem[]{} Lee, W. \& Ramirez-Ruiz, E. 2007, New J. Phys., 9, 17 

\bibitem[]{}  Li, L.-X., \& 
Paczy{\'n}ski, B.\ 1998, \apjl, 507, L59

\bibitem[]{} Lithwick, Y., \& Sari, R.\ 2001, \apj, 555, 540 

\bibitem[]{} Lyutikov, M., \& Blandford, R.\ 2003, arXiv:astro-ph/0312347 

\bibitem[]{} Lyutikov, M., \& Blackman, E.~G.\ 2001, \mnras, 321, 177 

\bibitem[]{} MacFadyen, A.~I., \& Woosley, S.~E.\ 1999, \apj, 524, 262 

\bibitem[MacFadyen et al.(2001)]{2001ApJ...550..410M} MacFadyen, A.~I., 
Woosley, S.~E., \& Heger, A.\ 2001, \apj, 550, 410 

\bibitem[]{} Manchester, R.~N.\ 2004, Science, 304, 542 

\bibitem[]{} Mazets, E.~P., et al.\ 1981, \apss, 80, 3 

\bibitem[]{} McKinney, J.~C.\ 2005, ArXiv 
Astrophysics e-prints, arXiv:astro-ph/0506368 

\bibitem[McKinney(2006)]{2006MNRAS.368.1561M} McKinney, J.~C.\ 2006, 
\mnras, 368, 1561 

\bibitem[McKinney 
\& Blandford(2009)]{2009MNRAS.394L.126M} McKinney, J.~C., \& Blandford, R.~D.\ 2009, \mnras, 394, L126 

\bibitem[Metzger et al.(2008)]{2008MNRAS.385.1455M} Metzger, B.~D., 
Quataert, E., \& Thompson, T.~A.\ 2008, \mnras, 385, 1455 

\bibitem[Metzger et al.(2007)]{2007ApJ...659..561M} Metzger, B.~D., 
Thompson, T.~A., \& Quataert, E.\ 2007, \apj, 659, 561 

\bibitem[Mezzacappa et al.(2001)]{2001PhRvL..86.1935M} Mezzacappa, A., 
Liebend{\"o}rfer, M., Messer, O.~E., Hix, W.~R., Thielemann, F.-K., 
\& Bruenn, S.~W.\ 2001, Physical Review Letters, 86, 1935 

\bibitem[Modjaz et al.(2006)]{2006ApJ...645L..21M} Modjaz, M., et al.\ 
2006, \apjl, 645, L21 

\bibitem[Morsony et al.(2007)]{2007ApJ...665..569M} Morsony, B.~J., 
Lazzati, D., \& Begelman, M.~C.\ 2007, \apj, 665, 569 

\bibitem[Murphy 
\& Burrows(2008)]{2008ApJ...688.1159M} Murphy, J.~W., \& Burrows, A.\ 2008, \apj, 688, 1159 

\bibitem[]{} Muno, M.~P., et al.\ 2006, \apjl, 636, L41 

\bibitem[]{} Nakar, E.\ 2007, PhR, 442, 166

\bibitem[Narayan 
\& Kumar(2009)]{2009MNRAS.394L.117N} Narayan, R., \& Kumar, P.\ 2009, \mnras, 394, L117  

\bibitem[Panaitescu(2008)]{2008MNRAS.383.1143P} Panaitescu, A.\ 2008, 
\mnras, 383, 1143 

\bibitem[]{} Piran, T.\ 2005, Reviews of 
Modern Physics, 76, 1143 

\bibitem[]{} Pons, J.~A., Reddy, S., 
Prakash, M., Lattimer, J.~M., \& Miralles, J.~A.\ 1999, \apj, 513, 780 

\bibitem[]{} Proga, D., \& Begelman, M.~C.\ 2003, \apj, 592, 767 

\bibitem[]{} Qian, Y.-Z., \& Woosley, S.~E.\ 1996, \apj, 471, 331 

\bibitem[]{} Rhoads, J.~E.\ 1997, \apjl, 487, L1

\bibitem[]{} Ruderman, M.\ 1975, New York 
Academy Sciences Annals, 262, 164 

\bibitem[Scheck et 
al.(2006)]{2006A&A...457..963S} Scheck, L., Kifonidis, K., Janka, H.-T., Muller, E.\ 2006, \aap, 457, 963 

\bibitem[]{} Smith, N., \& Owocki, S.~P.\ 2006, \apjl, 645, L45 

\bibitem[Soderberg et al.(2006)]{2006ApJ...638..930S} Soderberg, A.~M., 
Nakar, E., Berger, E., \& Kulkarni, S.~R.\ 2006, \apj, 638, 930 

\bibitem[]{} Spitkovsky, A.\ 2006, \apjl, 648, L51

\bibitem[Spruit(2004)]{2004BaltA..13..266S} Spruit, H.~C.\ 2004, Baltic 
Astronomy, 13, 266 

\bibitem[]{} Stanek, K.~Z., et al.\ 
2003, \apjl, 591, L17

\bibitem[Symbalisty(1984)]{1984ApJ...285..729S} Symbalisty, E.~M.~D.\ 1984, 
\apj, 285, 729 
 
\bibitem[]{} Taylor, G.~B., Frail, D.~A., Berger, E., \& Kulkarni, S.~R.\ 2004, \apjl, 609, L1 

\bibitem[Tchekhovskoy et al.(2008)]{2008MNRAS.388..551T} Tchekhovskoy, A., 
McKinney, J.~C., \& Narayan, R.\ 2008, \mnras, 388, 551 

\bibitem[Thompson(2006)]{2006ApJ...651..333T} Thompson, C.\ 2006, \apj, 
651, 333 

\bibitem[Thompson(2008)]{2008AIPC.1000..399T} Thompson, T.~A.\ 2008, 
American Institute of Physics Conference Series, 1000, 399 

\bibitem[]{} Thompson, T.~A., 
Chang, P., \& Quataert, E.\ 2004, \apj, 611, 380 

\bibitem[]{} Uhm, Z.~L., \& Beloborodov, A.~M.\ 2007, \apjl, 665, L93 

\bibitem[]{} Usov, V.~V.\ 1992, \nat, 357, 472

\bibitem[Uzdensky \& MacFadyen(2006)]{2006ApJ...647.1192U} Uzdensky, D.~A., \& MacFadyen, A.~I.\ 2006, \apj, 647, 1192 

\bibitem[]{} Weber, E.~J., \& Davis, L.~J.\ 1967, \apj, 148, 217 

\bibitem[]{} Wheeler, J.~C., Yi, I., H{\"o}flich, P., \& Wang, L.\ 2000, \apj, 537, 810 

\bibitem[]{} Woods, P.~M., \& Thompson, C.\ 2006, Compact stellar X-ray sources, 547 

\bibitem[]{} Woosley, S.~E.\ 1993, \apj, 405, 273 

\bibitem[]{} Woosley, S.~E., \& Bloom, J.~S.\ 2006, Annual Rev.~of A$\&$A, 44, 507 

\bibitem[Woosley 
\& Heger(2006)]{2006ApJ...637..914W} Woosley, S.~E., \& Heger, A.\ 2006, \apj, 637, 914 

\bibitem[]{} Woosley, S.~E., \& Weaver, T.~A.\ 1995, \apjs, 101, 181 

\bibitem[Zalamea 
\& Beloborodov(2009)]{2009MNRAS.398.2005Z} Zalamea, I., \& Beloborodov, A.~M.\ 2009, \mnras, 398, 2005 

\bibitem[]{} Zhang, B., \& M{\'e}sz{\'a}ros, P.\ 2004, International Journal of Modern Physics A, 19, 2385 

\bibitem[Zhang et al.(2004)]{2004ApJ...608..365Z} Zhang, W., Woosley, 
S.~E., \& Heger, A.\ 2004, \apj, 608, 365 

\end{thebibliography}
\end{document}